\documentclass[prl,twocolumn,showpacs,preprintnumbers,amsmath,amssymb,floats]{revtex4-1}
\usepackage{graphicx}
\usepackage{amsmath}
\usepackage{epstopdf}
\usepackage{amsbsy}
\usepackage{color}
\usepackage{dcolumn}
\usepackage{bm}

\begin{document}

\title{Origin of electronic dimers in the spin-density wave phase of Fe-based superconductors}
\author{Maria N. Gastiasoro$^1$, P. J. Hirschfeld$^2$, and Brian M. Andersen$^1$}
\affiliation{$^1$Niels Bohr Institute, University of Copenhagen, Universitetsparken 5, DK-2100 Copenhagen,
Denmark\\
$^2$Department of Physics, University of Florida, Gainesville, Florida 32611, USA}

\date{\today}

\begin{abstract}

We investigate the emergent impurity-induced states arising from point-like scatterers in the spin-density wave phase of iron-based superconductors within a microscopic five-band model. Independent of the details of the band-structure and disorder potential, it is shown how stable magnetic $(\pi,\pi)$ unidirectional nematogens are formed locally by the impurities. Interestingly, these nematogens exhibit a dimer structure in the electronic density, are directed along the antiferromagnetic $a$-axis, and have typical lengths of $\sim10$ lattice constants in excellent agreement with recent scanning tunnelling experiments. These electronic dimers provide a natural explanation of the dopant-induced transport anisotropy found e.g. in the 122 iron pnictides.

\end{abstract}


\maketitle

Hints of electronic nematicity, i.e.  spontaneous breaking of discrete rotational symmetry while preserving translational symmetry, appear in many cases among the Fe-based materials\cite{hu12,fernandes12}. These anisotropies, detected by STM\cite{chuang10}, transport (on detwinned samples)\cite{tanatar,chu,chu2}, ARPES\cite{kim11,yi11,yi12,zhang12}, neutron scattering\cite{zhao}, optical\cite{nakajima} and Raman\cite{gallais} spectroscopy, and torque magnetometry\cite{kasahara}, have largely been interpreted in terms of an {\it intrinsic} tendency of these systems to break $C_4$ symmetry globally due to nematic correlations, due either to spin nematic effects\cite{hu12,fernandes12} or orbital ordering.\cite{kruger09,chen09,lv09,lee09} However, there are also indications of local defect states which break $C_4$ symmetry\cite{grothe12,song11,song12,hanaguri12,zhou11,allan13}.   Since impurities are known to nucleate magnetic order locally\cite{alloul09,gastiasoro13,gastiasoroprl13},
it is reasonable to assume that incipient nematic order in a fluctuating state can be similarly condensed around a defect, to create a
local nematic electronic state.  Once present, such anisotropic
defect structures can influence macroscopic anisotropy as well, and it has been suggested\cite{chuang10,allan13,ishida13,rosenthal13} that they are responsible for the resistivity anisotropy    observed in detwinned Ba-122 crystals\cite{chu,chu2,ishida13}.

Because impurities represent a well-defined perturbation which can be examined locally,  studying the electronic states they create can yield important information on the background correlations present in the pure system\cite{alloul09}.  In YBCO, for example, magnetic droplets formed around
Zn impurities are known to have a size consistent with the pure antiferromagnetic (AF) correlation length.\cite{alloul09}
At present, the microscopic mechanism  responsible for the creation of local defect states in Fe-based materials and for breaking of rotational symmetry is
 unclear.  Some clues are offered by STM experiments on defects in the underdoped, magnetically ordered phase, where the symmetry is already broken
   by the ($\pi,0$) spin-density wave (SDW) order (the wave vector is given in the effective 1-Fe Brillouin zone).   Fourier transform scanning tunneling spectroscopy (STS)
   deduced the existence of electronic defects with $C_2$ symmetry\cite{chuang10} nucleated by the Co dopants in Ca(Fe$_{1-x}$Co$_x$)$_2$As$_2$, while a more detailed analysis reported a dimer structure.\cite{allan13} These dimers are approximately eight lattice constants ($a$) long and oriented along the AF $a$-axis, consistent with the larger resistivity found along the ferromagnetic $b$-axis.

As a first step in understanding the origin of local $C_4$ symmetry breaking observed in several experiments on various materials, it
 seems useful to study a situation where a known chemical impurity substitutes at a known position, and ask why such
 a dimer-like structure (with charge or local density of states (LDOS) peaks located such a great distance from the impurity site) should be
 induced.  As in the cuprates, this problem should be accessible to weak-coupling theories of these systems, provided they  account for the
electronic states of the system to which the impurity couples and treat interactions on the average.   Hints that an unusual electronic state might
be induced were found already in first principles calculations of a Co dopant in Ba-122, where the magnetic potential due to the Co was found
to be oscillatory and exceed several unit cells\cite{kemper}. Impurity-induced $C_2$ structures have been previously studied within a strong-coupling model\cite{chen_NJP}, and a scenario based on a competing pocket density wave order which has, however, not been observed\cite{kang}. Finally, in Ref. \onlinecite{kontani}, impurities were shown to pin fluctuating orbital order and create local states with broken $C_4$ symmetry but no dimer-like character.

Here, we present a general study the origin of electronic dimer states  by an explicit, unbiased microscopic five-band calculation of the  local electronic densities near a point-like impurity potential in the SDW phase of the iron pnictides. The impurity causes a local $(\pi,\pi)$ magnetic instability, which combined with the $(\pi,0)$ order of the bulk SDW phase, results in unidirectional magnetic defects oriented along the AF $a$-axis, and associated electronic density dimers. These dimers are caused by local density modulations of the SDW phase, and not dependent on the details of the band-structure, disorder potential, or local impurity-modified interaction parameters. This is a concrete example of a phenomenon which is largely
unexplored, the local nucleation of a particular magnetic order in a bulk
state with different magnetic order. We show how such emergent impurity states evolve from droplets at high temperatures $T$ to nematogens in the low-$T$ SDW phase. The final size of the low-$T$ dimers is consistent with recent STM measurements\cite{allan13}, but depends within our theory on the "cooling rate", and we show how dimers of other lengths may also be obtained. Finally, we compute the LDOS characteristics of the dimer states to compare with STM experiments, and discuss how our model can be used to perform
realistic  calculations of effective defect potentials for application to transport experiments.

The five-orbital Hamiltonian consists of three terms
\begin{equation}
 \label{eq:H}
 H=H_{0}+H_{int}+H_{imp},
\end{equation}
with $H_0$  the kinetic part arising from a  tight-binding fit to the density functional theory (DFT) band-structure of Ref.~\onlinecite{graser} (similar results arise by using e.g. the band of Ikeda {\it et al.}\cite{ikeda10})
\begin{equation}
 \label{eq:H0}
H_{0}=\sum_{\mathbf{ij},\mu\nu,\sigma}t_{\mathbf{ij}}^{\mu\nu}c_{\mathbf{i}\mu\sigma}^{\dagger}c_{\mathbf{j}\nu\sigma}-\mu_0\sum_{\mathbf{i}\mu\sigma}n_{\mathbf{i}\mu.\sigma}.
\end{equation}
The operators $c_{\mathbf{i} \mu\sigma}^{\dagger}$ ($c_{\mathbf{i}\mu\sigma}$) create (annihilate) an electron at site $i$ in orbital state $\mu$ with spin $\sigma$, and $\mu_0$ is the chemical potential fixed such that the doping $\delta=\langle n \rangle - 6.0$ is zero.
The indices $\mu$ and $\nu$ denote  the five iron orbitals $d_{xy}$, $d_{xz}$, $d_{yz}$, $d_{x^2-y^2}$, and $d_{3z^2-r^2}$. The second term in Eq.(\ref{eq:H}) describes the onsite Coulomb interaction
\begin{align}
 \label{eq:Hint}
 H_{int}&=U\sum_{\mathbf{i},\mu}n_{\mathbf{i}\mu\uparrow}n_{\mathbf{i}\mu\downarrow}+(U'-\frac{J}{2})\sum_{\mathbf{i},\mu<\nu,\sigma\sigma'}n_{\mathbf{i}\mu\sigma}n_{\mathbf{i}\nu\sigma'}\\\nonumber
&\quad-2J\sum_{\mathbf{i},\mu<\nu}\vec{S}_{\mathbf{i}\mu}\cdot\vec{S}_{\mathbf{i}\nu}+J'\sum_{\mathbf{i},\mu<\nu,\sigma}c_{\mathbf{i}\mu\sigma}^{\dagger}c_{\mathbf{i}\mu\bar{\sigma}}^{\dagger}c_{\mathbf{i}\nu\bar{\sigma}}c_{\mathbf{i}\nu\sigma},
\end{align}
which includes the intraorbital (interorbital) interaction $U$ ($U'$), the Hund's rule coupling $J$ and the pair hopping energy $J'$.
We assume spin rotation invariance $U'=U-2J$ and $J'=J$. In this work we additionally take $J=U/4$. The last term in Eq.\eqref{eq:H} describes the impurity
\begin{equation}
 H_{imp}=V_{imp}\sum_{\mu\sigma}c_{\mathbf{i^*}\mu\sigma}^{\dagger}c_{\mathbf{i^*}\mu\sigma},
\end{equation}
which adds a local potential $V_{imp}$ at the impurity site $\mathbf{i^*}$. In the present work, we neglect the orbital dependence of the impurity potential for simplicity.

After a mean-field decoupling of Eq.\eqref{eq:Hint} we solve the following eigenvalue problem $\sum_{\mathbf{j}\nu}
H^{\mu\nu}_{\mathbf{i} \mathbf{j} \sigma}
 u_{\mathbf{j}\nu}^{n}
=E_{n} u_{\mathbf{i}\mu}^{n}$,
where
\begin{align}
 H^{\mu\nu}_{\mathbf{i} \mathbf{j} \sigma}&=t_{\mathbf{ij}}^{\mu\nu}+\delta_{\mathbf{ij}}\delta_{\mu\nu}[-\mu_0+\delta_{\mathbf{ii^*}}V_{imp}+U \langle n_{\mathbf{i}\mu\bar{\sigma}}\rangle\\\nonumber
\quad&+\sum_{\mu' \neq \mu}(U'\langle n_{\mathbf{i}\mu' \bar{\sigma}}\rangle+(U'-J)\langle n_{\mathbf{i}\mu' \sigma}\rangle)],
 \end{align}
on $N_x\times N_y$ lattices with self-consistently obtained $\langle n_{\mathbf{i}\mu\sigma} \rangle=\sum_{n}|u_{\mathbf{i}\mu\sigma}^{n}|^{2}f(E_{n\sigma})$,  with $f$  the Fermi function. We note that while several calculations of inhomogeneous states have been performed using realistic, five-orbital models\cite{kaneshita09,luo10,bascones10,ikeda10,brydon11,lv11,schmiedt12}, none have discussed the role of residual electronic correlations and induced local SDW order.

\begin{figure}[t]
\begin{center}
\includegraphics[width=4.35cm]{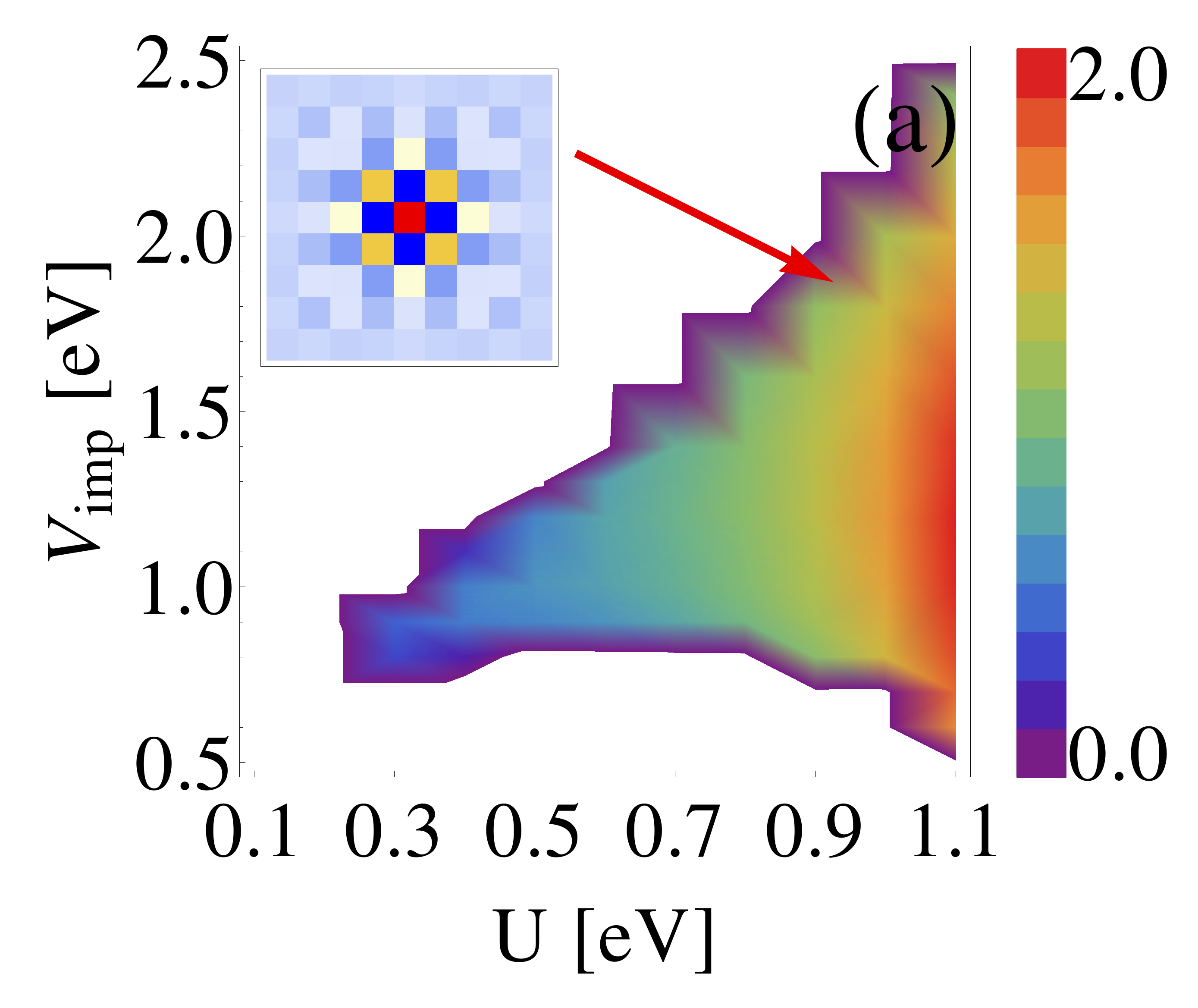}
\includegraphics[width=4.2cm]{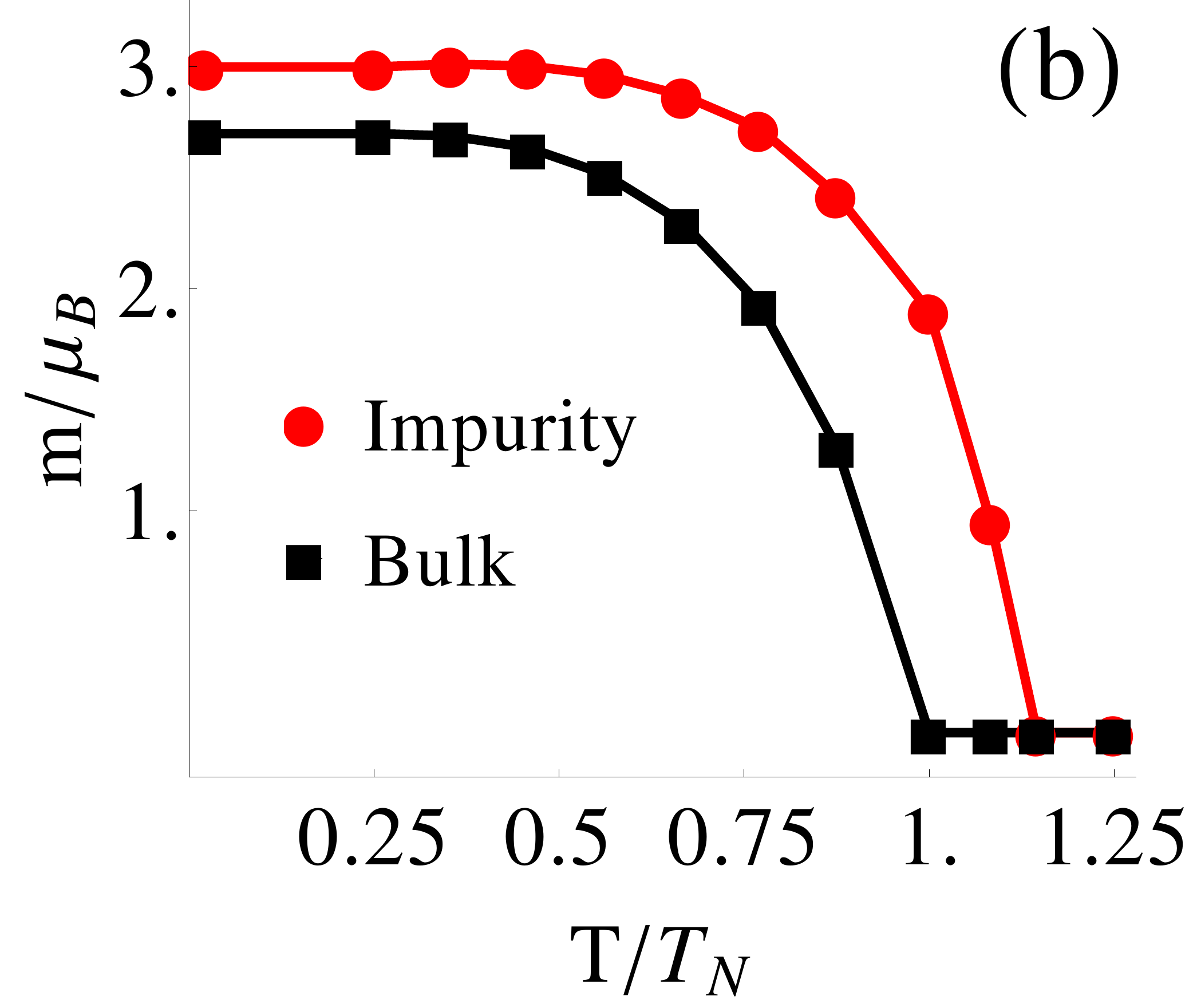}
\end{center}
\caption{(Color online) (a) Single impurity phase diagram displaying the impurity-induced magnetization at the impurity site at $T=0$  vs. $U$
and  $V_{imp}$. Inset: spatial magnetization of the induced impurity state at $T=0$ in the paramagnetic state. (b) Magnetization vs. $T$ (normalized to the bulk N\'{e}el temperature $T_N$) at the impurity site (red curve) and in the bulk (black) with $U=1.6$eV ($T_N=0.3U$), and $V_{imp}=0.1$eV.}
\label{fig:1}
\end{figure}

In the absence of disorder, the ground state of the undoped system is a normal metal or a metallic $(\pi,0)$ SDW depending on the value of $U$. However, even without SDW order,  impurities can pin magnetic order locally\cite{alloul09,gastiasoro13,gastiasoroprl13} similar to what has been discussed extensively for cuprate superconductors.\cite{tsuchiura01,wang02,zhu02,chen04,kontani06,harter07,andersen07} In Fig.~\ref{fig:1}(a) we show the single impurity phase diagram at low $T$ for impurity-induced magnetic order vs. $U$ and $V_{imp}$ (below the bulk SDW phase transition at $U_{c2}\simeq 1.2$eV for $T=0$). As seen, the potential generates local commensurate $(\pi,\pi)$ magnetic order in a regime of intermediate strength repulsive $V_{imp}$ for sufficiently large $U$.
 
\begin{figure}[t]
\begin{center}
\includegraphics[width=7.65cm]{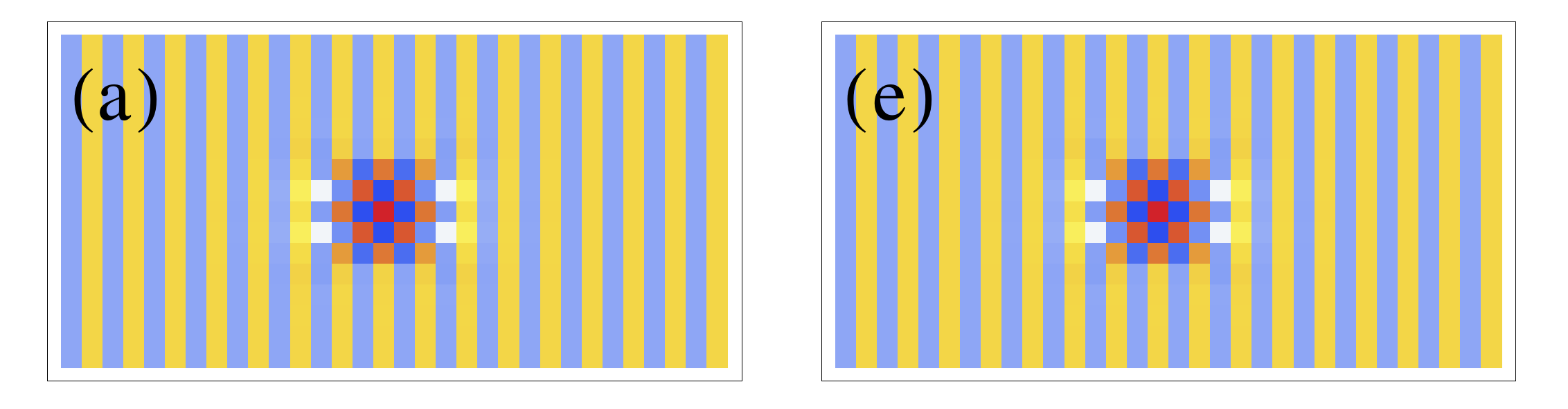}
\includegraphics[width=0.87cm]{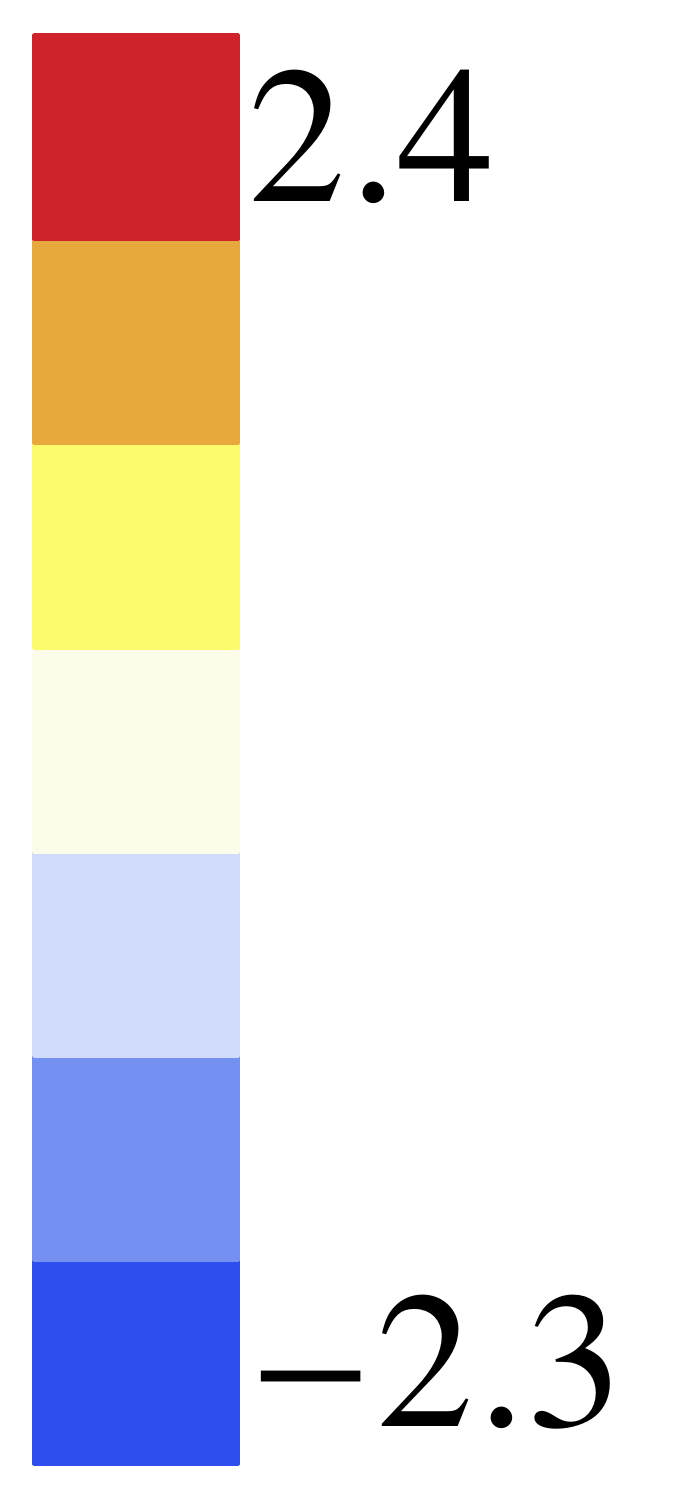}
\\
\includegraphics[width=7.65cm]{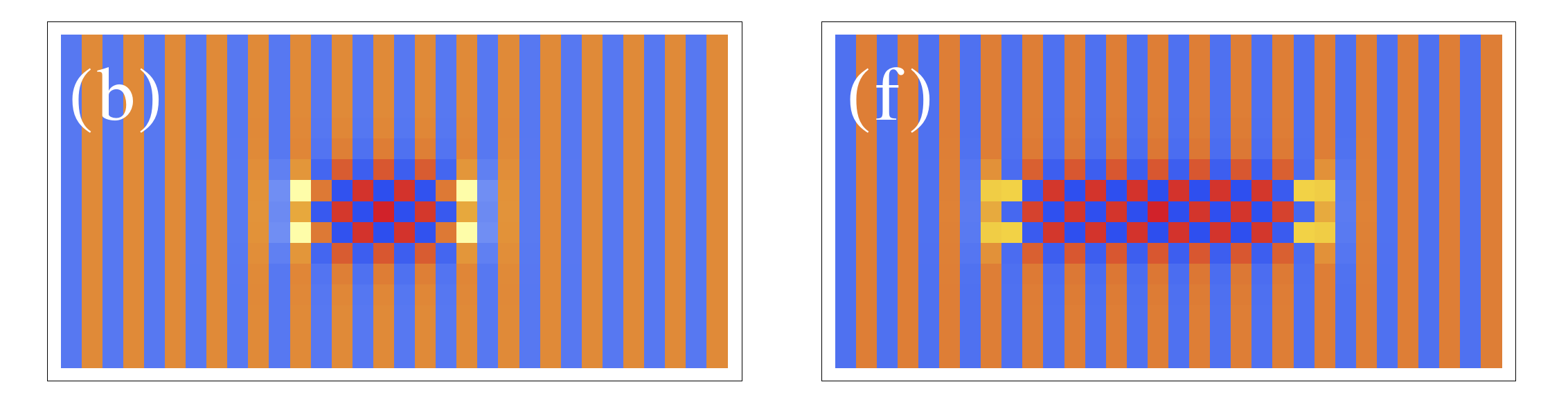}
\includegraphics[width=0.87cm]{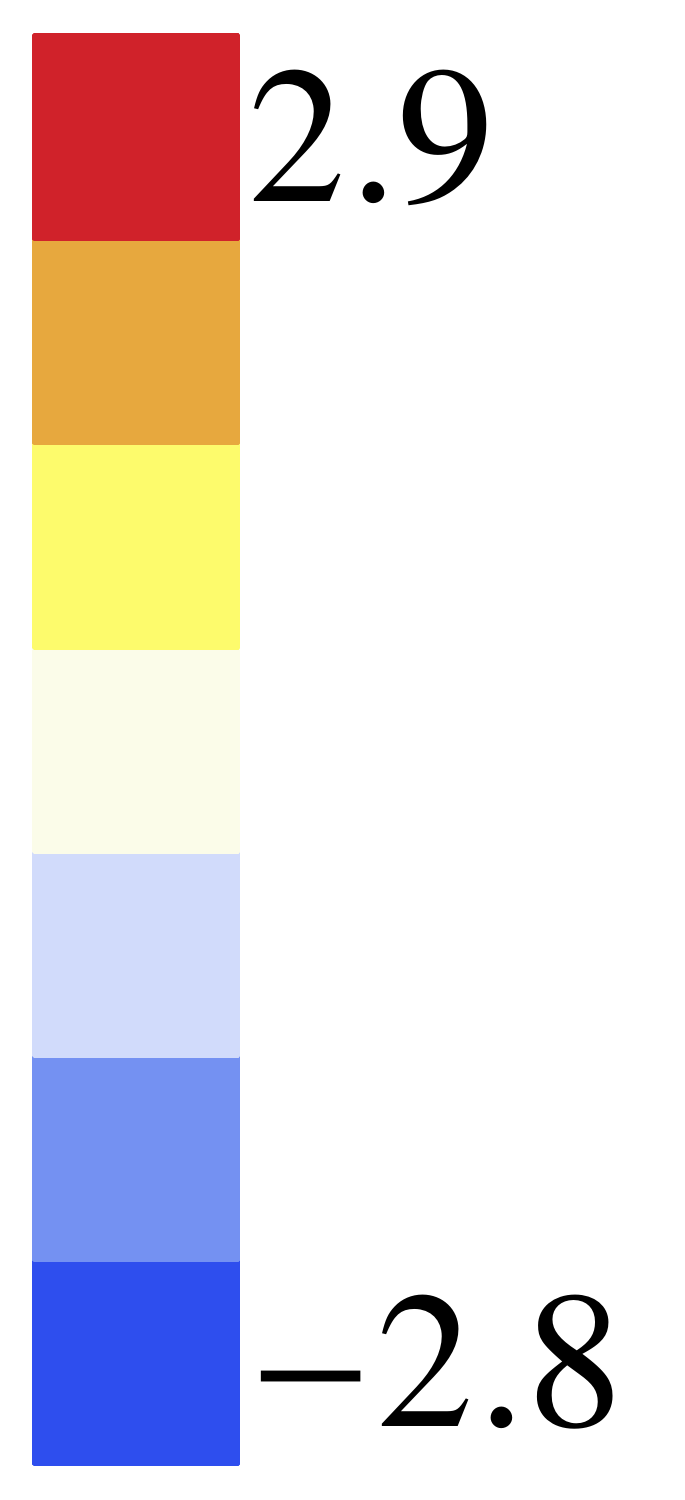}
\\
\includegraphics[width=7.65cm]{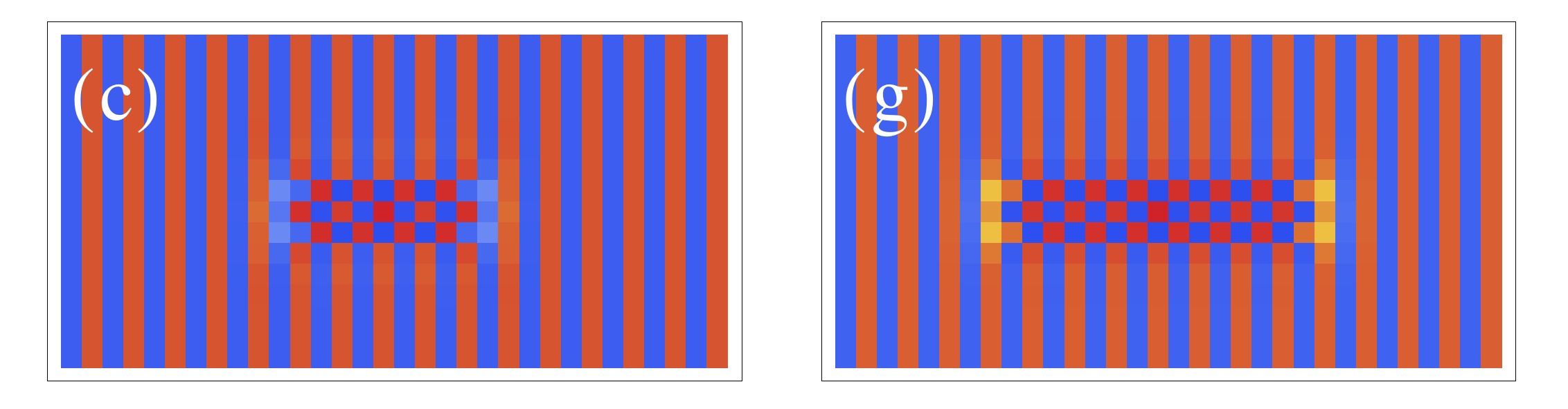}
\includegraphics[width=0.87cm]{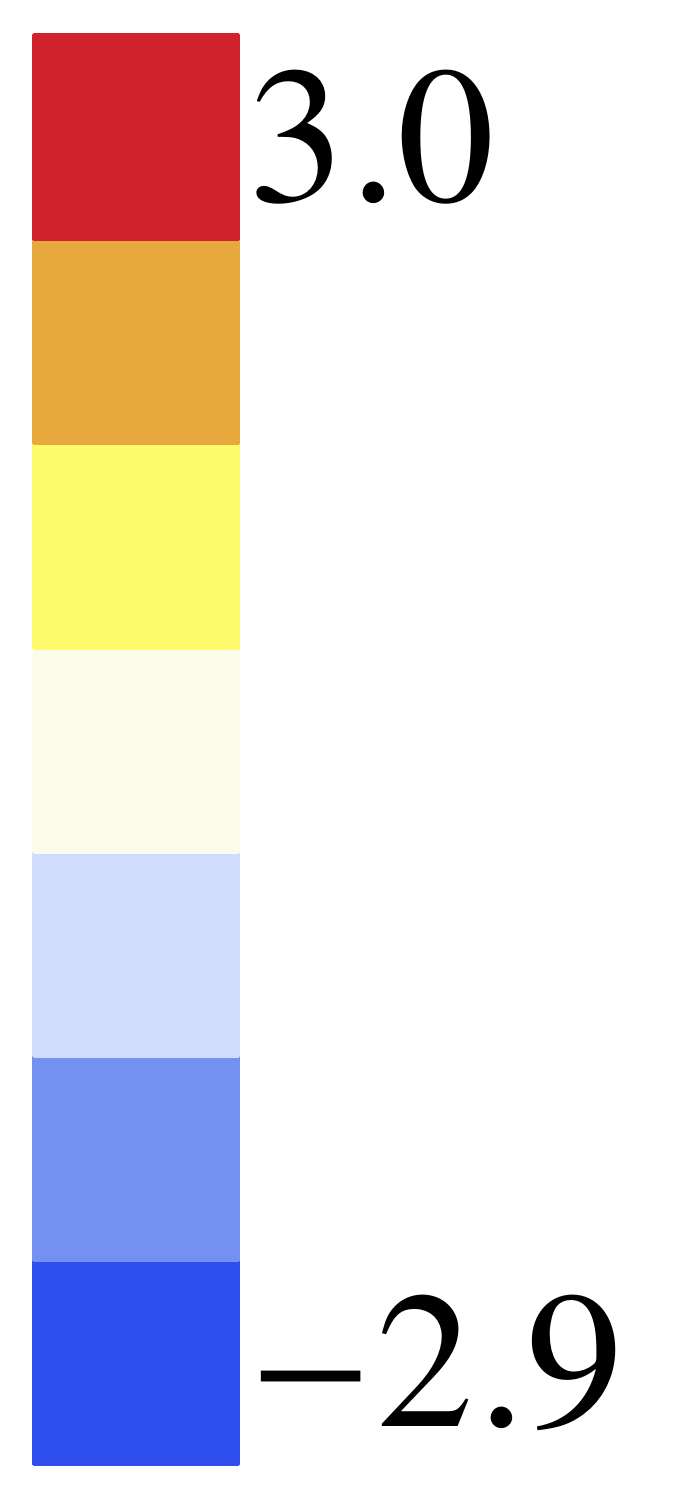}
\\
\includegraphics[width=7.65cm]{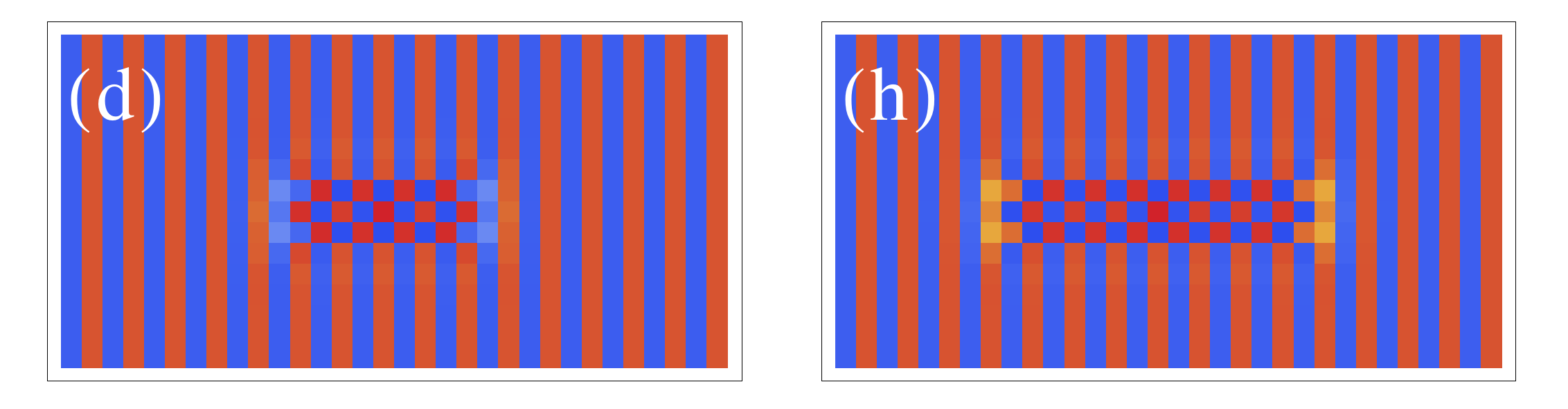}
\includegraphics[width=0.87cm]{leg22.pdf}
\end{center}
\caption{(Color online) Magnetization in real-space upon lowering $T$ with $U=1.6$eV and $V_{imp}=0.1$eV. From top to bottom: $T/T_N=0.88, 0.67, 0.46, 0.25$. At each $T$, the system is iterated until the bulk magnetization has converged. The final low-$T$ nematogen is stable and fully converged, but its length depends on the "cooling rate", as seen by comparison of the left and right columns distinguished by different cooling steps $\Delta T/T_N$=0.21 (left), 0.105 (right). Note that only half of the steps are shown for the slow cooling case.}
\label{fig:2}
\end{figure}

When $U$ exceeds $U_{c2}$, similar impurity-induced order takes place at high $T$ above $T_N$ as shown in Figs.~\ref{fig:2}(a,e). The local magnetic moment at the potential site is displayed in Fig.~\ref{fig:1}(b) where one clearly sees the extended magnetic impurity phase above $T_N$, and the enhanced impurity moments at low $T$ over the bulk SDW magnetization.
Upon lowering $T$, the $C_4$ symmetric high-$T$ magnetic ($\pi,\pi$) droplets shown in Fig.~\ref{fig:2}(a,e) have to compete with the surrounding bulk ($\pi,0$) SDW order. As shown in Fig.~\ref{fig:2}, the $C_2$ structure of the SDW phase leads to a magnetic cigar-like impurity structure, a nematogen aligned along the AF $a$-axis, which still exhibits the internal ($\pi,\pi$) magnetic structure of the high-$T$ phase, but inherits an overall $C_2$ symmetry from the SDW background. While these low-$T$ nematogens shown in Fig.~\ref{fig:2} are stable and fully converged, their final length depends on the path of convergence, i.e. the number of steps taken in the cooling process, as seen explicitly by comparing the right and left columns in Fig.~\ref{fig:2}. The origin of this "cooling rate" dependence is simply a competition between the impurity-induced ($\pi,\pi$) moments and the SDW long-range order. It is natural to speculate that this  numerical cooling-rate dependence is related to the significant annealing dependence of the resistivity anisotropy found recently.\cite{nakajima,ishida13} We note that the existence of the nematogens is robust, and not dependent on the cooling rate, band-structure, or correlation strength $U$ (as long as $U>U_{c2}$), or strength of the impurity potential $V_{imp}$, provided the latter two are able to create a high-$T$ impurity bound state [Fig.~\ref{fig:2}(a,e)].

\begin{figure}[b]
\begin{center}
\includegraphics[width=7.65cm]{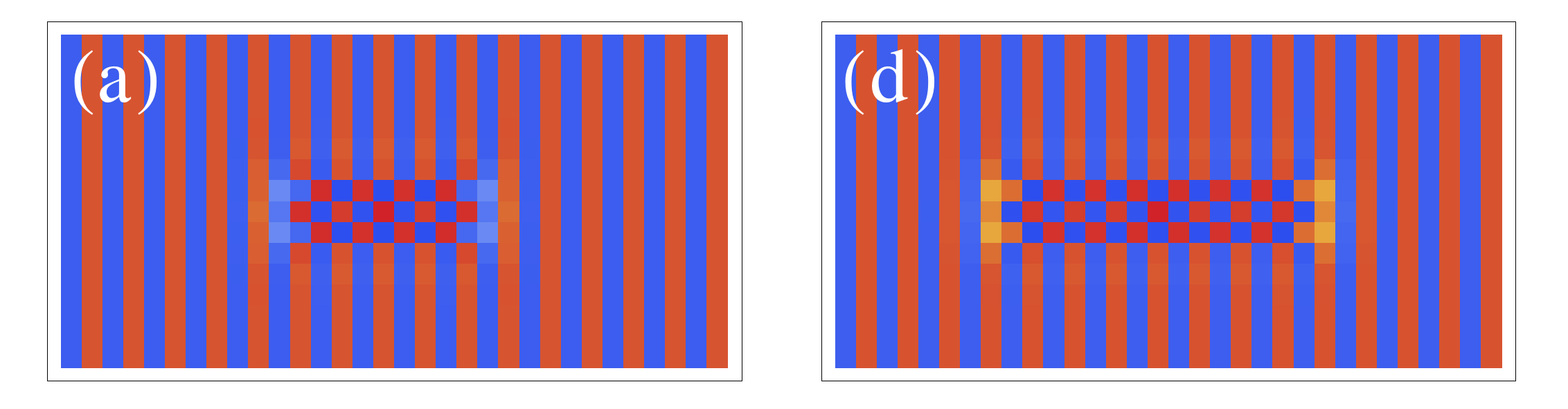}
\includegraphics[width=0.87cm]{leg22.pdf}
\\
\includegraphics[width=7.65cm]{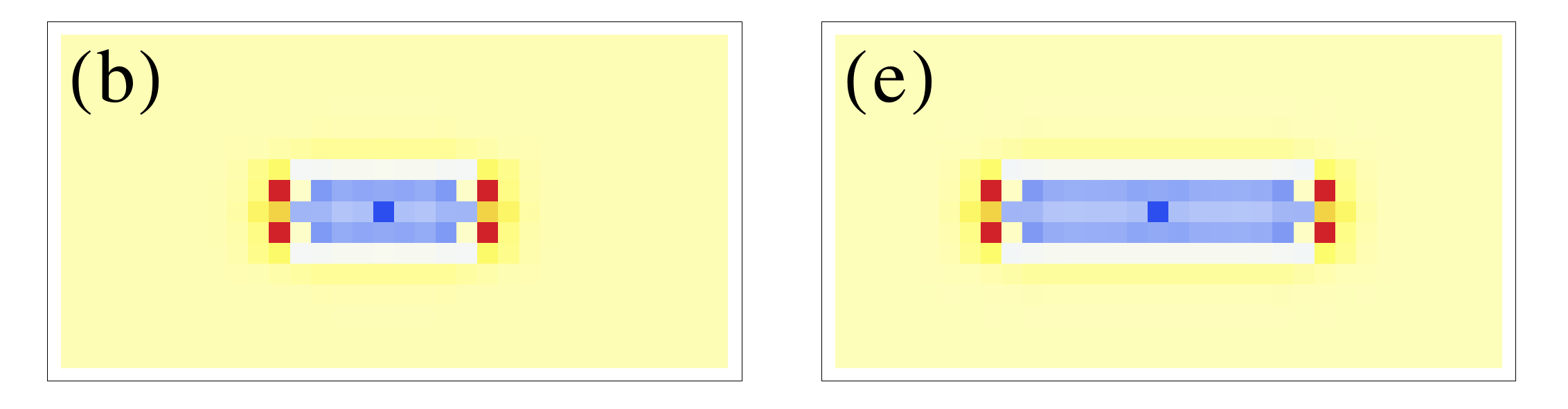}
\includegraphics[width=0.87cm]{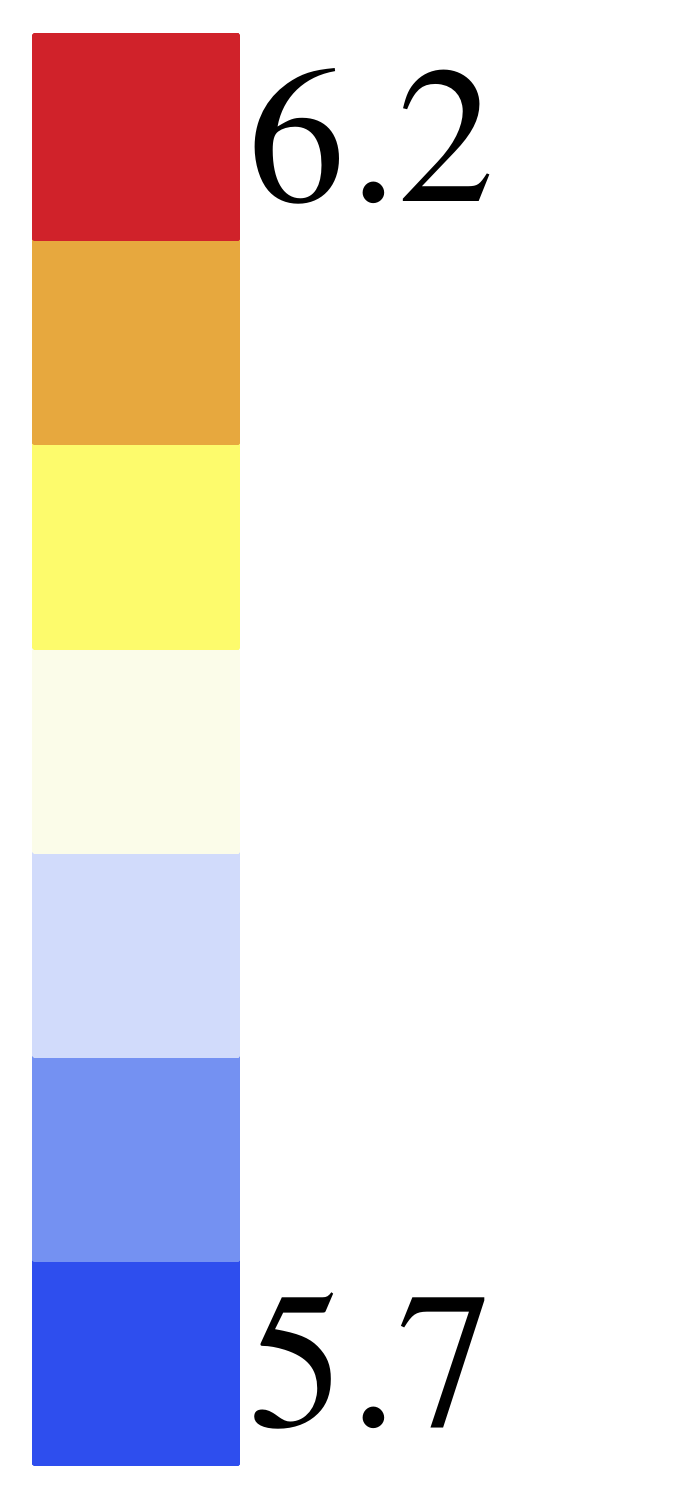}
\\
\includegraphics[width=7.65cm]{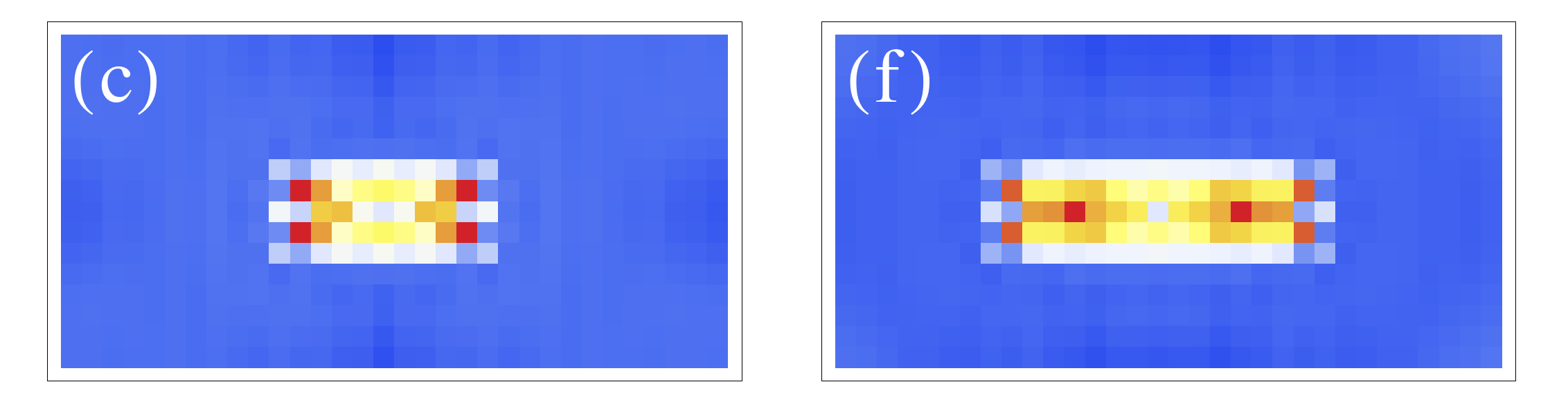}
\includegraphics[width=0.87cm]{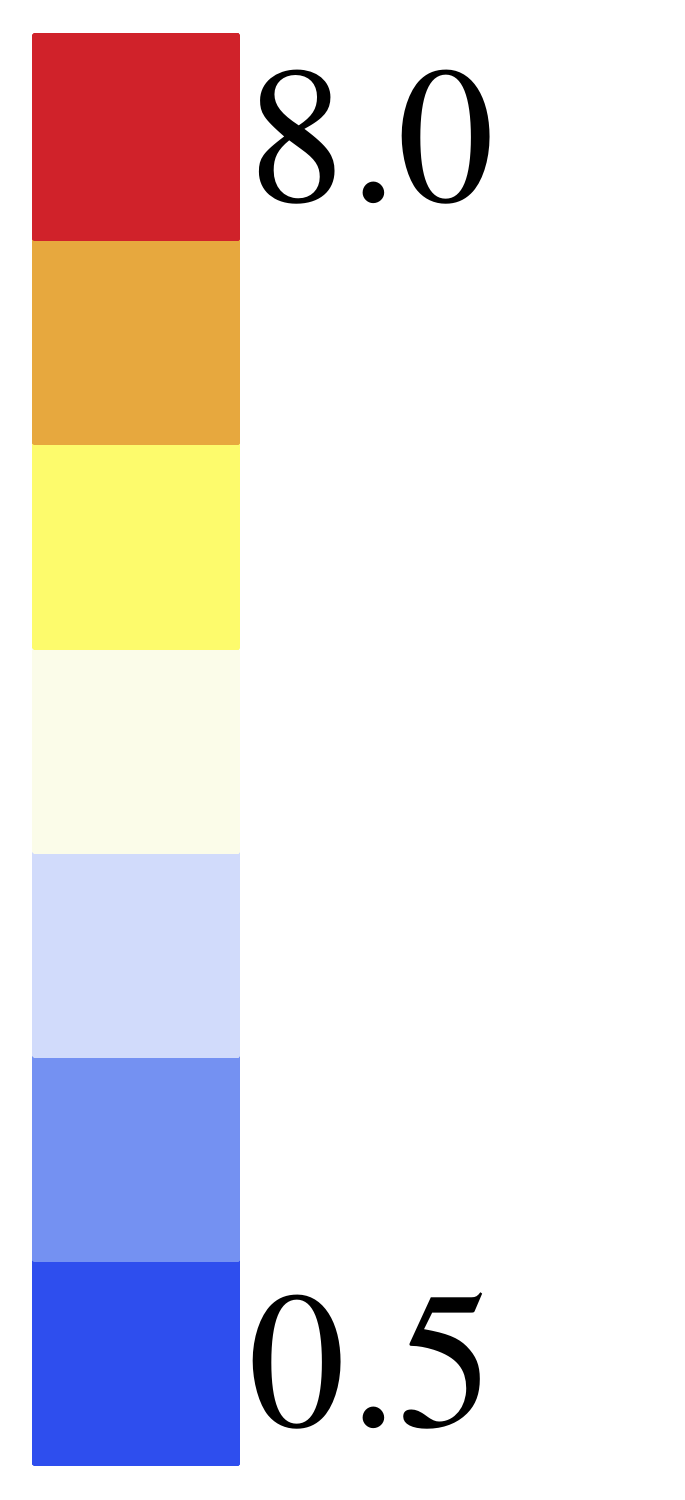}
\end{center}
\caption{(Color online) 2D real-space maps of (a,d) the magnetization, (b,e) the total electron charge density, and (c,f) the low-energy integrated LDOS for the same two low-$T$ nematogens shown in Fig.~\ref{fig:2}(d,h).}
\label{fig:3}
\end{figure}

Figure~\ref{fig:3} summarizes the magnetic and electronic properties of the unidirectional low-$T$ nematogens. As seen from Fig.~\ref{fig:3}(a,d), the impurity-induced magnetization nematogen consists of an odd number of sites (even number of lattice spacings) in length and width, and is always oriented along the AF
$a$-axis.  This is clearly because such a structure can lower
its energy with a long unfrustrated boundary with the $(\pi,0)$
order.  In addition, in Fig.~\ref{fig:3}(b,e) we see that the nematogen exhibits peaks at
both ends resulting in an electronic dimer-like structure
of the total charge density. This appears to be a general
characteristic of these emergent impurity states, and also
follows from the same energetic considerations, since the
magnetic energy can be lowered by creating a charge state as homogeneous as possible; thus the excess charge 
from the impurity site is moved to the ends of the nematogen. Previous STM work has discovered the existence of electronic dimers in the LDOS,\cite{grothe12,song11,song12,hanaguri12} and recent partially integrated LDOS within the SDW phase found strong evidence for electronic dimers near Co dopants in CaFe$_2$As$_2$.\cite{allan13} Within the present scenario, the existence of density dimers naturally explains the presence of LDOS dimers. However, as opposed to the robust existence of the nematogens, the detailed structure of the LDOS near the impurity sites is more sensitive to parameters. In Fig.~\ref{fig:3} we show cases where the onsite impurity potential is weak, $V_{imp}=0.1$eV, and the integrated LDOS from -37meV to 0meV, which is identical to the range used in Ref. \onlinecite{allan13}, indeed produces an LDOS dimer of $\sim8a$ in both cases. 

\begin{figure}[t]
\begin{center}
\includegraphics[width=0.6cm]{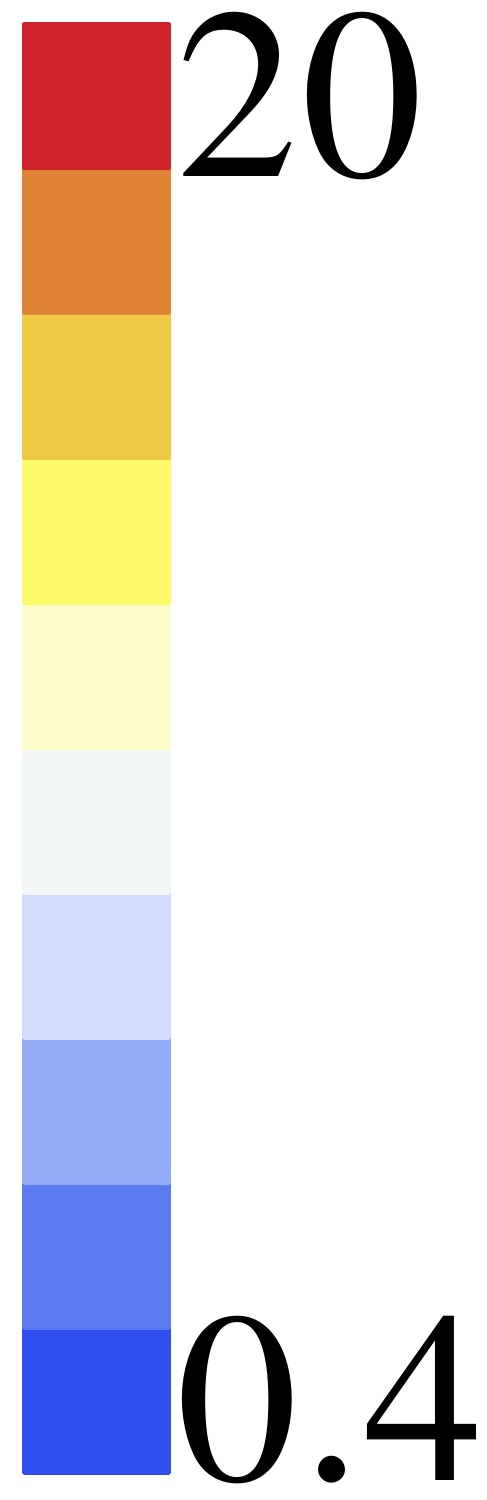}
\includegraphics[width=7.95cm]{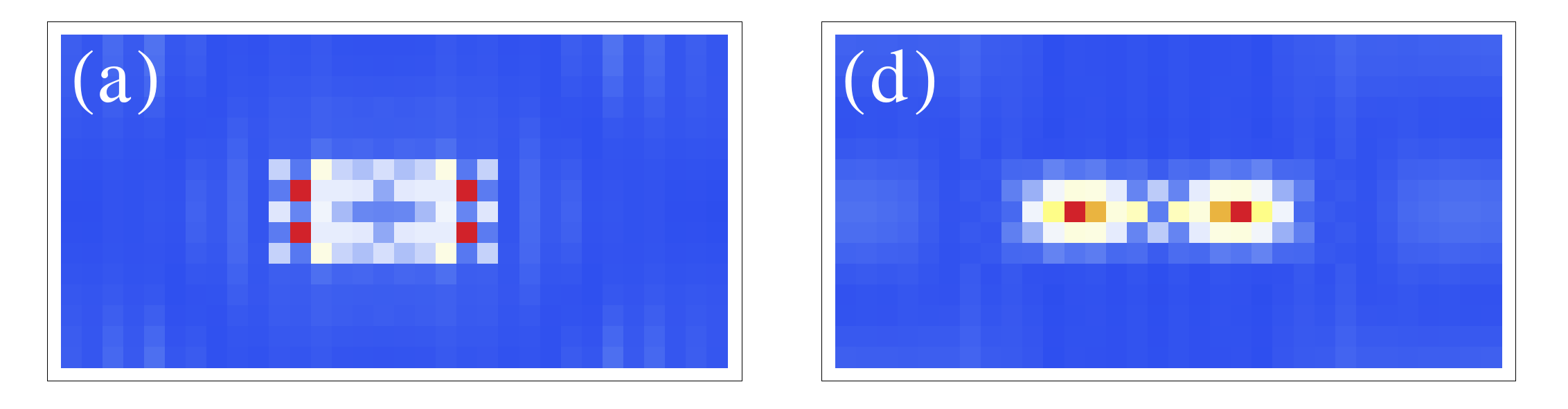}
\\
\includegraphics[width=0.6cm]{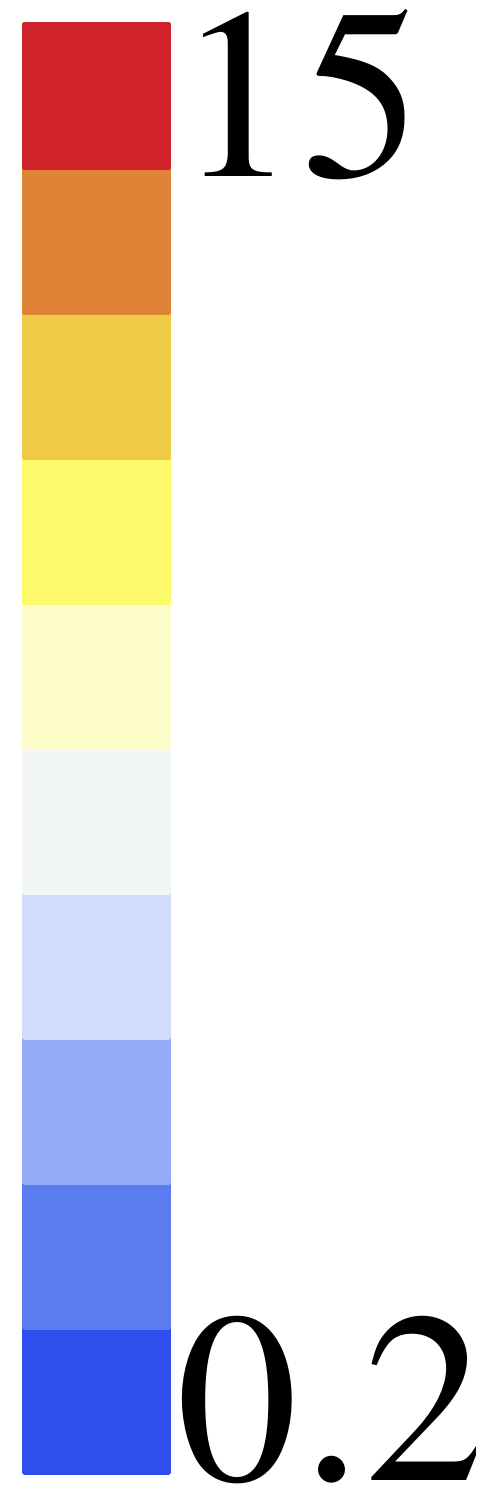}
\includegraphics[width=7.95cm]{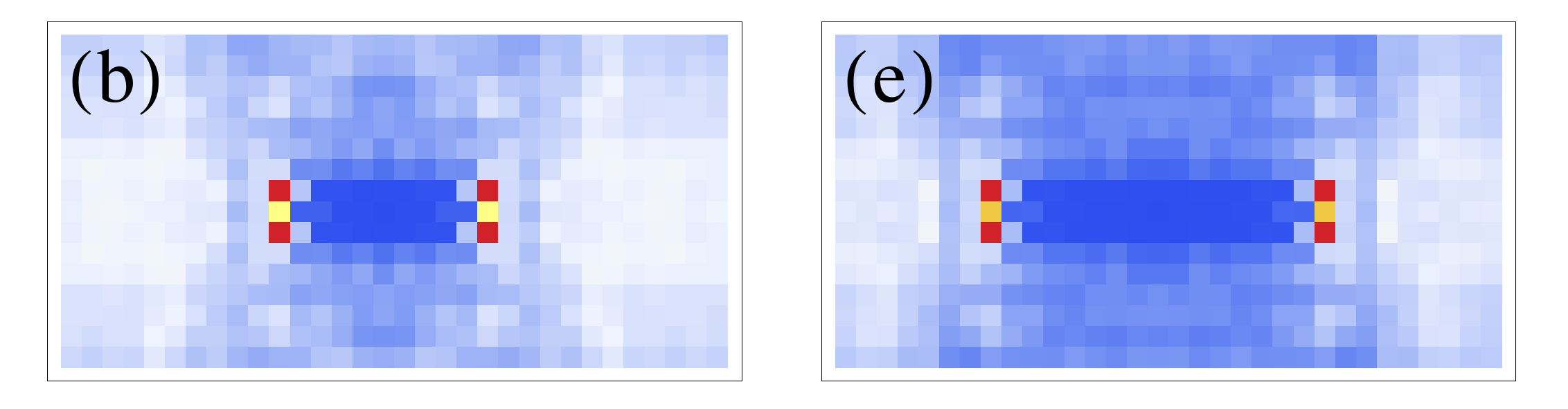}
\\
\includegraphics[width=8.5cm]{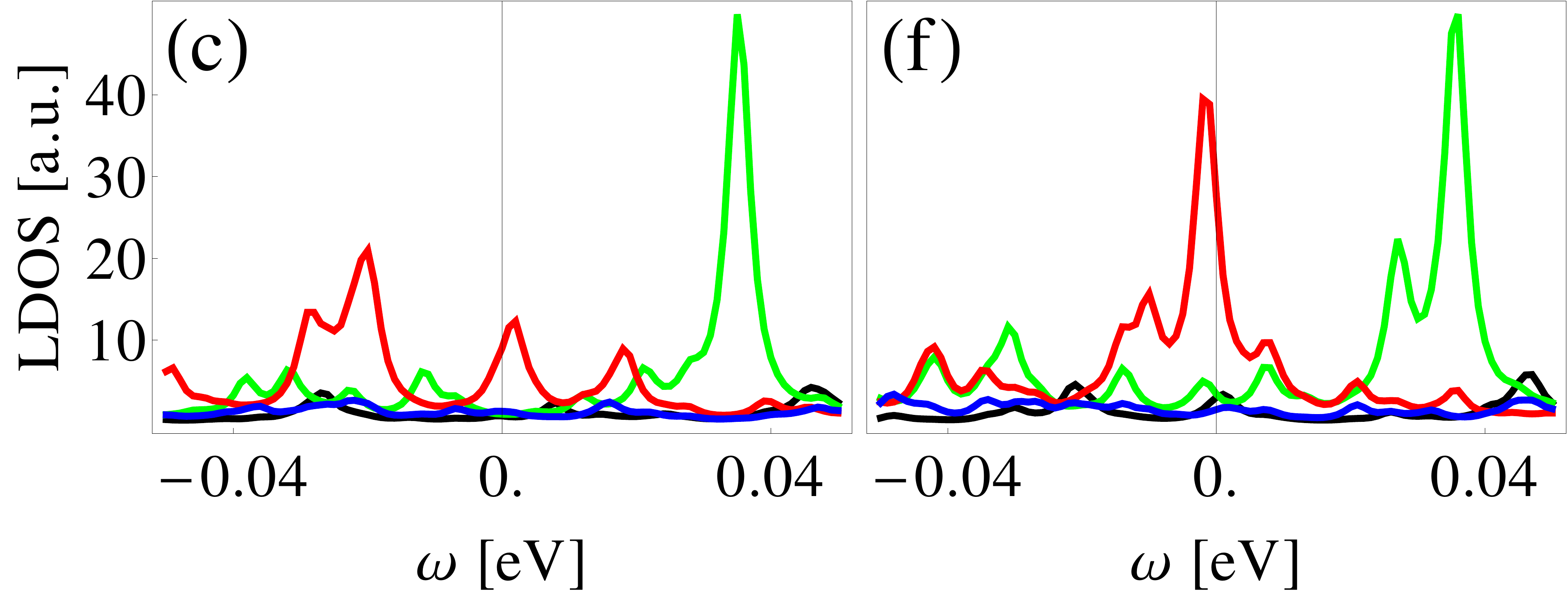}
\end{center}
\caption{(Color online) (a,b,d,e) 2D real-space maps of the LDOS at the characteristic low-energy dimer energy $\omega=-20$meV (a) and $\omega=-2$meV (d) and at the charge peak energy $\omega=-330$meV (b,e) for the nematogens in Fig.~\ref{fig:3}(b,e). Panels (c,f) display LDOS vs. $\omega$ for different sites in the system: bulk (black), impurity site (green), at the low-energy dimer peak (red), and at the charge density peak (blue).}
\label{fig:4}
\end{figure}

Figure~\ref{fig:4} displays in greater detail the LDOS properties of the nematogens. The first point we wish to illustrate is that the size of the dimers deduced from the low-energy integrated LDOS is not necessarily the same as that in the (fully integrated) charge distribution. The low-energy integrated LDOS shown in Figs.~\ref{fig:3}(c,f) exhibits the dimer structure because of a peak in the LDOS at a particular site within the low-energy integration window. Figures~\ref{fig:4}(a,d) show the LDOS at an energy corresponding to the peak in the LDOS at this site, yielding essentially the same result as in Figs.~\ref{fig:3}(c,f). For the current parameters, the LDOS at the (distinct) sites where the charge distribution is maximal [Fig.~\ref{fig:3} (b,e)] exhibits a peak  at higher binding energy,  as shown in Figs.~\ref{fig:4}(b,e). The structure of the low-energy LDOS can be more clearly inferred from Figs.~\ref{fig:4}(c,f) which show the energy dependence of the LDOS at several relevant sites in the nematogen. From the LDOS on the site corresponding to the low-energy dimer peak,
for example, it is evident why the low-energy integrated LDOS exhibits a peak at roughly $4a$ from the impurity site in this case. 

For the LDOS results presented here we 
have used a potential strength of $V_{imp}=0.1$eV for all orbitals since the low-energy LDOS results in a dimer similar to the STM results in Ref.~\onlinecite{allan13}.
First principles studies have found that in the normal state (SDW state) Co atoms in the FeAs plane correspond to attractive\cite{nakamura11} (repulsive\cite{kemper}) intra-orbital potentials. We have investigated both cases and obtained qualitatively the same results, i.e. nematogens and electronic density dimers though the details of the LDOS are sensitive to the host bandstructure and the value of the impurity potential, as expected. It will be interesting in future studies to combine the results presented here with more detailed models for the impurity potentials to obtain  quantitative agreement with STS measurements on the dimers for specific impurities and hosts.

One interesting feature of the current simulation is that, while all orbitals contribute roughly equally to the magnetic order of the nematogen shown in Fig.~\ref{fig:3}(a,d), only certain components of the impurity potential appear to be important.  In the current case, we find that mainly the $d_{3z^2-r^2}$ component is important for inducing the impurity bound state and magnetic structure at both low-$T$ [Fig.~\ref{fig:1}(a)] and high-$T$ [Fig.~\ref{fig:2}(a,e)]. The LDOS curves in Figs.~\ref{fig:4}(a-d) are also mainly dominated by the d$_{3z^2-r^2}$ orbital. This is somewhat unusual since the Fermi surface and the bulk SDW magnetization is dominated by the t$_{2g}$ orbitals. We believe that the importance of the $d_{3z^2-r^2}$ orbital arises from the fact that the impurity potential allows for mixing with the $d_{3z^2-r^2}$-dominated band which exhibits a sharp peak in the DOS about 0.25 eV below the Fermi level for the bandstructure used in this work.\cite{graser}  

Local $C_4$ symmetry breaking defect signatures, including vortex states\cite{song11}, have been observed in many different Fe based
superconductors.   We have not exhibited a mechanism for explaining all of them, but have begun the process of realistic modeling of those which
can be expected to be driven by bulk magnetic order.   The remarkable existence of large scale charge dimers induced around the impurities in our calculations is consistent with
STM measurements on Ca-122\cite{allan13}, but it is intriguing that even larger electronic dimers have also been imaged by STM in FeSe samples which are not thought
to be magnetic\cite{song11,song12}.  This may imply that the thin film FeSe samples used in STM in fact display at least intermediate range magnetic order which is difficult to detect\cite{feng13},
or possibly that dimers can be created in a strongly fluctuating paramagnetic phase if the $C_4$ symmetry is already broken.  The latter possibility   is an important question to address in the context of anisotropic transport measurements on Ba-122\cite{chu,chu2,ishida13}, where
significant anisotropy enhancement is observed below the structural phase transition $T_s$ but above $T_N$.  While the simulations we have presented here are consistent with the sign and magnitude of the transport anisotropy in the ordered phase $T<T_N$, with the current tetragonal bandstructure we find that impurity-induced nematogens do not form above $T_N$.    Whether they are stable in the structural symmetry broken phase $T_N<T<T_s$ will be the subject of a
future study.

In summary, we have discovered the emergence of unidirectional electronic dimer states induced by point-like defects in the SDW phase of iron pnictides. These dimers arise from  a competition between the impurity-induced ($\pi,\pi$) magnetism and the surrounding bulk SDW ($\pi,0$) phase. They are oriented along the AF $a$-axis and exhibit typical lengths of $\sim10a$ in agreement with STM measurements. While they occur robustly with realistic bandstructures for iron-based materials, we have also shown that e.g. their size and LDOS characteristics are not universal, and discussed the factors which govern these properties. These nematogens are excellent microscopic candidates for the origin of at least part of the transport anisotropy observed in detwinned crystals.

We thank M. Allan, S. Mukherjee, and I. Paul for useful discussions. B.M.A. and M.N.G. acknowledge support from the Lundbeckfond fellowship (grant A9318). P.J.H. was supported by NSF-DMR-1005625.

\end{document}